# An Integral Field Spectrograph for SNAP Supernova Identification

A. Ealet[m]; E. Prieto[m]; A. Bonissent[g]; R. Malina[m]; G. Bernstein[e], S. Basa[m]; O. LeFevre[m], A. Mazure[m], C. Bonneville[m];C. Akerlof[b], G. Aldering[a], R. Amanullah[c], P. Astier[d], E. Barrelet[d], C. Bebek[a], L. Bergström[c], J. Bercovitz[a], M. Bester[f], C. Bower[h],W. Carithers[a], E. Commins[f], C. Day[a], S. Deustua[i], R. DiGennaro[a], R. Ellis[j],M. Eriksson[c], A. Fruchter[k], J-F. Genat[d], G. Goldhaber[f], A. Goobar[c], D. Groom[a], S. Harris[f], P. Harvey[f], H. Heetderks[f], S. Holland[a], D. Huterer[l], A. Karcher[a], A. Kim[a],W. Kolbe[a], B. Krieger[a], R. Lafever[a], J. Lamoureux[a], M. Lampton[f], M. Levi[a], D. Levin[b],E. Linder[a], S. Loken[a], R. Massey[n], T. McKay[b], S. McKee[b], R. Miquel[a], E. Mörtsell[c], N. Mostek[h], S. Mufson[h], J. Musser[h], P. Nugent[a], H. Oluseyi[a], R. Pain[d], N. Palaio[a], D. Pankow[f], S. Perlmutter[a], R. Pratt[f], A. Refregier[n], J. Rhodes[o], K. Robinson[a], N. Roe[a], M. Sholl[f], M. Schubnell[b], G. Smadja[p], G. Smoot[f], A. Spadafora[a], G. Tarle[b], A. Tomasch[b], H. von der Lippe[a], D. Vincent[d], J-P. Walder[a], G. Wang[a]

[a] Lawrence Berkeley National Laboratory, Berkeley CA, USA
[b] University of Michigan, Ann Arbor MI, USA
[c] University of Stockholm, Stockholm, Sweden
[d] CNRS/IN2P3/LPNHE, Paris, France
[e] University of Pennsylvania, Philadelphia PA, USA
[f] University of California, Berkeley CA, USA
[g] CNRS/IN2P3/CPPM, Marseille, France
[h] Indiana University, Bloomington IN, USA
[i] American Astronomical Society, Washington DC, USA
[j] California Institute of Technology, Pasadena CA, USA
[k] Space Telescope Science Institute, Baltimore MD, USA
[l] Case Western Reserve University, Cleveland OH, USA
[m] CNRS/INSU/LAM, Marseille, France
[n] Cambridge University, Cambridge, UK
[o] NASA Goddard Space Flight Center, Greenbelt MD, USA
[p] CNRS/IN2P3/INPL, Lyon, France

**Contact**:
Anne.ealet@oamp.fr, tel +33 4 91 05 59 00, Laboratoire d'Astrophysique de Marseille, BP 8, 13376 Marseille, France

## ABSTRACT

A well-adapted spectrograph concept has been developed for the SNAP (SuperNova/Acceleration Probe) experiment. The goal is to ensure proper identification of Type Ia supernovae and to standardize the magnitude of each candidate by determining explosion parameters. An instrument based on an integral field method with the powerful concept of imager slicing has been designed and is presented in this paper. The spectrograph concept is optimized to have very high efficiency and low spectral resolution (R~100), constant through the wavelength range (0.35-1.7µm), adapted to the scientific goals of the mission.

## 1. INTRODUCTION

The SNAP satellite is designed to measure very precisely the cosmological parameters and to determine the nature of the dark energy. The mission is based on the measurement of some 2000 supernovae (SNe) of Type Ia up to a redshift of $z$=1.7. Details of the mission and the expected physics results are provided in other contributions[0,2,34,5] to these Proceedings. Spectroscopy of each candidate supernova near maximum light is required to (a) insure that only Type Ia supernovae are used, and (b) measure the physical parameters of the supernova explosion through various spectral features. The parameters of the explosion, *e.g.* the progenitor metallicity, have small but important effects upon the peak magnitude and color of Type Ia events, and hence must be measured to insure the highest possible accuracy in determination of the cosmological parameters.

## 2. SCIENCE DRIVERS

To achieve the primary goals of selecting Type Ia supernovae and of controlling for intrinsic physical variations, a spectrum of each candidate is must be acquired near maximum light. Operating in space has clear advantages for such a mission:

- Easy access to the infrared region
- A very stable and small PSF
- 24h operation
- Lower exposure time, thanks to reduction of background to the level of the zodiacal light

In this case the major limitation is the aperture of the primary mirror (2m) and the need for the best overall efficiency to access the faintest supernovae (magnitude ~25 at peak).

A typical supernova spectrum of Type Ia at peak is shown in Figure 1. Of particular interest is the breadth of all lines, which indicates that high-resolution spectroscopy is not required. The specific signature of Type Ia supernovae is the SiII line at $\lambda$=6150Å (rest frame). This line is very broad (~200Å rest frame) and is broadened further by the redshift factor (1+$z$) and shifted to ~1.7µm for $z$=1.7. Other types of SNe, such as II or Ib, have lines of H or He in the same wavelength range, allowing the classification of all possible candidates. The feature characteristics (position, width, height, *etc*.) are directly related to the peak magnitude through physical parameters such as temperature, velocity and progenitor metallicity. In models, the strongest sensitivity to the metallicity in the progenitor system lies in the rest-frame UV band, which defines a broad wavelength range, 0.4<$\lambda$<1.7 µm, that must be covered by the instrument. The broad features of the SNe spectra and the non-negligible detector noise contribution for the faintest objects make a low-resolution spectrograph optimal: a resolving power $\lambda/\delta\lambda$~100 at FWHM and 1 pixel per FWHM sampling, with constant resolving power in the 0.6-1.7µm range, is required to keep the S/N optimized for all redshifts. The main specifications are summarized in Table 1.

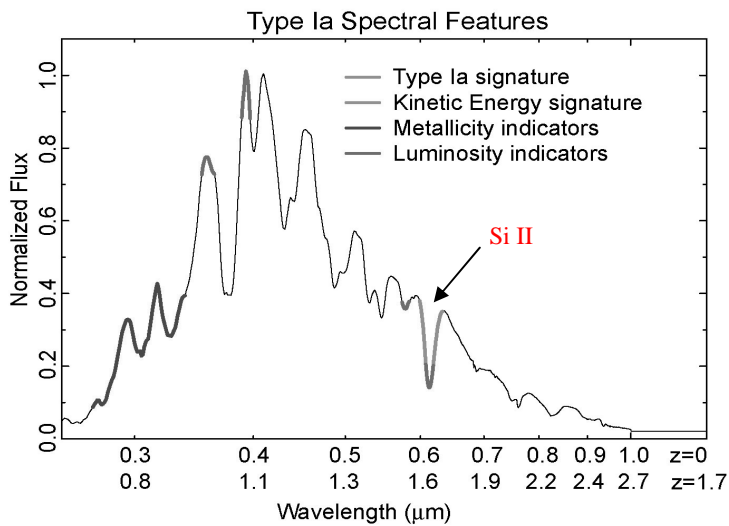

Figure 1: Typical spectrum of SN Ia.

| Property | Visible | IR |
|---|---|---|
| Wavelength coverage (μm) | 0.35-0.98 | 0.98-1.70 |
| Field of view | 3.0" × 3.0" | 3.0" × 3.0" |
| Spatial resolution element (arc sec) | 0.15 | 0.15 |
| Number of slices | 20 | 20 |
| Spectral resolution, $\lambda/\delta\lambda$ | 100 | 100 |

Table 1: Spectrograph main specifications.

The field of view should include the underlying galaxy in order to determine its spectrum during the same exposure. This is necessary for subtraction of the host spectrum from the spectrum in the supernovae region. Based on the mean size of galaxies of redshift 1-2, a 3"×3" field of view is sufficient.

Space instruments must be light and compact, and must minimize interfaces with the spacecraft. To reduce the time budget, we must take the galaxy and SN spectra in the same exposure and push to the highest efficiency possible for the highest redshifts.

## 3. INSTRUMENT CONCEPT TRADE-OFF

Given the science drivers and requirements, we conducted a trade-off study to choose the best instrument concept. The requirement for simultaneous acquisition of SN and host spectra, and the high object acquisition precision that would be needed for a traditional long slit spectrograph, lead us to prefer a 3D spectrograph. A 3D spectrograph reconstructs the data cube including the two spatial directions X and Y plus the wavelength direction as shown in Figure 2. For each spatial pixel, the spectrum is reconstructed. Thanks to the 3"×3" field of view, the pointing requirements are relaxed and the galaxy and SN data are acquired at the same time. Two principal techniques are indicated for 3D spectroscopy: first, the use of a Fourier Transform Spectrometer (FTS), and second, the use of integral field spectroscopy.

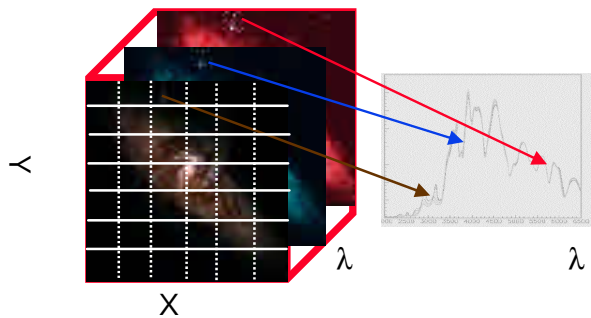

Figure 2: 3D spectroscopy illustration.

The FTS technique is based on the classical Michelson interferometer principle. When one of the two flat mirrors is moved, the Fourier transform space of the wavelength is scanned. Herschel is using this technique. But the domain of excellence for the FTS is at longer wavelengths, smaller wavelength range, and higher spectral resolution than SNAP needs. The main drawback to an FTS on SNAP would be the need for a translating device with a quarter-millimeter throw and a positioning accuracy of a few nanometers. This would call for a complex mechanism, even more so for a very precise metrology system.

Integral field spectroscopy using traditional dispersers is based on three generic techniques as shown in Figure 3. The simplest one is the use of

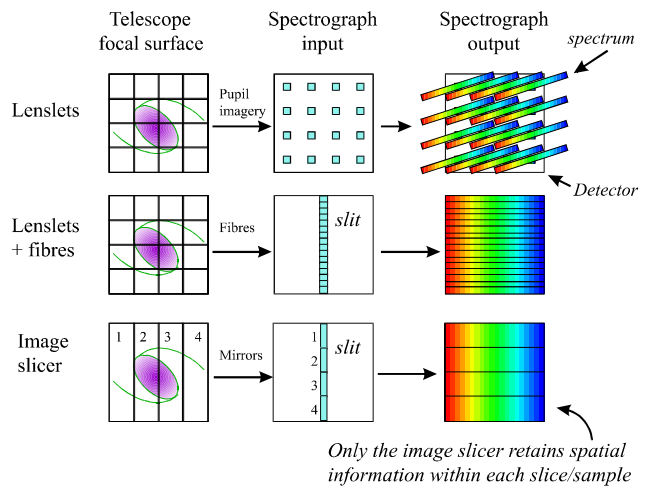

*Only the image slicer retains spatial information within each slice/sample*

Figure 3: Integral Field Spectroscopy illustration.

classical microlenses in order to create an image of the telescope pupil at the entrance of the spectrograph (Tigre-CFHT[7]). The dispersion takes place between the pupil images. As there is little room for each spectrum on the focal plane, a filter wheel clearly is needed. This is excluded for SNAP, where all information should be acquired in one shot.

In order to have room to accommodate the spectra, we must rebuild an entrance slit for the spectrograph. This would be possible using microlenses and fibers to convey the light (*i.e.* VIMOS[5,8] and GMOS). However, as the coupling efficiency between fibers and microlenses is not optimal, this solution has only an average efficiency. As SNAP is only a 2-meter-class telescope, instead we must treasure every photon.

The final possibility in our trade-off is the image slicer technique. This technique, developed since 1938 in order to minimize slit losses, is very powerful[5,9,10 11]. The new generation of image slicers improves the efficiency and the compactness of the system. Figure 4 shows the principle of this technique. The field of view is sliced along $N$ (in the drawing $N=3$, for SNAP $N=20$) strips on a called slicing mirror (stack of $N$ plates where the active surface is on an edge called a slice). Each of $N$ slices re-images the telescope pupil, creating $N$ telescope pupil images in the pupil plane. Thanks to a tilt adapted to each individual slice, the $N$ pupil images lie along a line. In the pupil plane, a line of so-called "pupil" mirrors is arranged. Each pupil mirror is placed on a pupil image and it re-images the field strip. These images are arranged along a line and form a "pseudo-slit." At this stage, therefore, we have an image of each of the $N$ strips of the field of view. The pseudo-slit is placed in the entrance plane of the spectrograph, acting as the entrance slit.

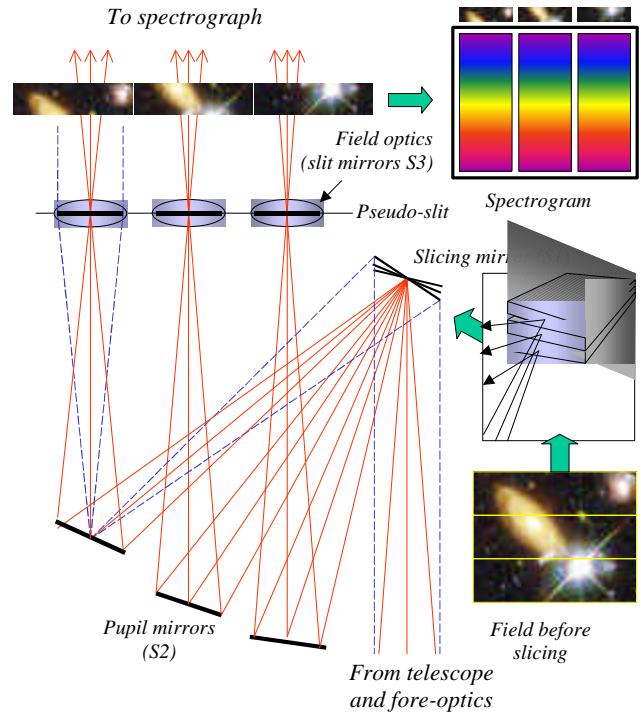

Figure 4. Image slicer principle (courtesy J. Allington-Smith, Durham U.)

A last line of mirrors is placed on the pseudo-slit. This line adapts the output pupil of the slicer into the input pupil of the spectrograph.

## 4. INSTRUMENT CONCEPT

The instrument functionalities are summarized in the block diagram shown in Figure 5. Principal components are described below.

**RELAY OPTICS**

This unit is the interface between the telescope beam and the instrument. The optical solution is highly dependent on the implementation of the instrument. The definition of this optical system requires knowledge of the spectrograph position with respect to the telescope focal plane. The beam can be picked off wherever it is most convenient for the overall instrument. It will be beneficial to correct some telescope aberrations within this optical system. A simple, easily conceived three-mirror configuration should be sufficient to satisfy these requirements.

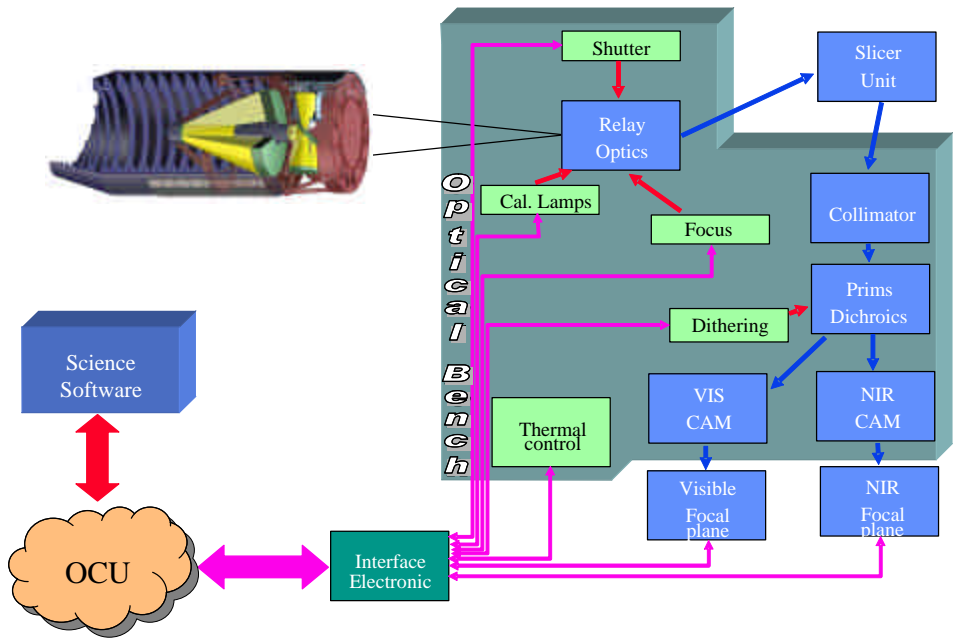

Figure 5: Instrument block diagram.

**SLICER UNIT**

The slicer unit acts as a field reformatting system. As described above, the principle is to slice a 2D field of view into long strips and optically align all the strips to form a long spectrograph entrance slit. The slicing mirror is comprised of a stack of slicers. Each slicer has an optically active spherical surface on one edge (see Figure 6). A line of pupil mirrors does the reformatting. Each pupil mirror sends the beam to a slit mirror, which adapts the pupil to the entrance of the spectrograph.

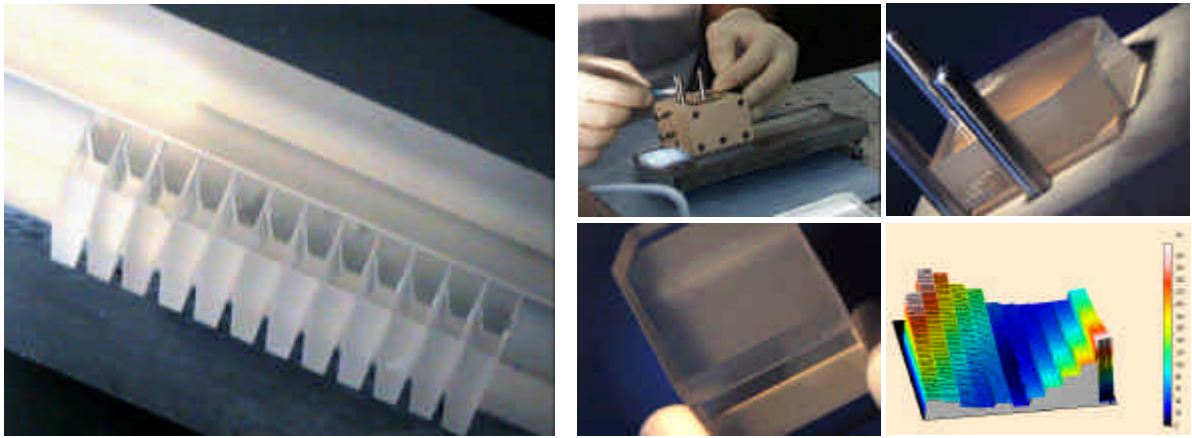

Figure 6: Left: line of pupil mirrors assembled; right: view of the complete slicer mirror.

The long thin active surface of each individual slicer will produce a large diffraction effect. In order to minimize flux losses to a few percent, the spectrograph entrance pupil must be oversized. A combined theoretical and experimental approach is underway at LAM to define the optimum entrance pupil (in the infrared bands 1-5 µm).

The baseline requirements on the slicer unit are an accuracy of $\leq \lambda/10$ rms on the optical surfaces and a surface roughness of $\leq 5$ nm rms. Existing prototypes fully meet these specifications.

**OPTICAL BENCH**

Thanks to the moderate beam aperture and field of view, the spectrograph optics will be straightforward. The baseline is a classical dichroic spectrograph: one collimator mirror, one prism with a dichroic crystal, and two camera mirrors are required. Using spherical shapes for all the mirrors would provide an adequately sharp image, but using aspheric mirrors will make it possible to have a very compact spectrograph. The prism solution is well matched to the requirement of a flat resolution over the whole wavelength range. The dichroic crystal allows covering two channels simultaneously: one for the visible (*e.g.*. 0.35-0.98 µm) and one for the infrared (0.98-1.70 µm).

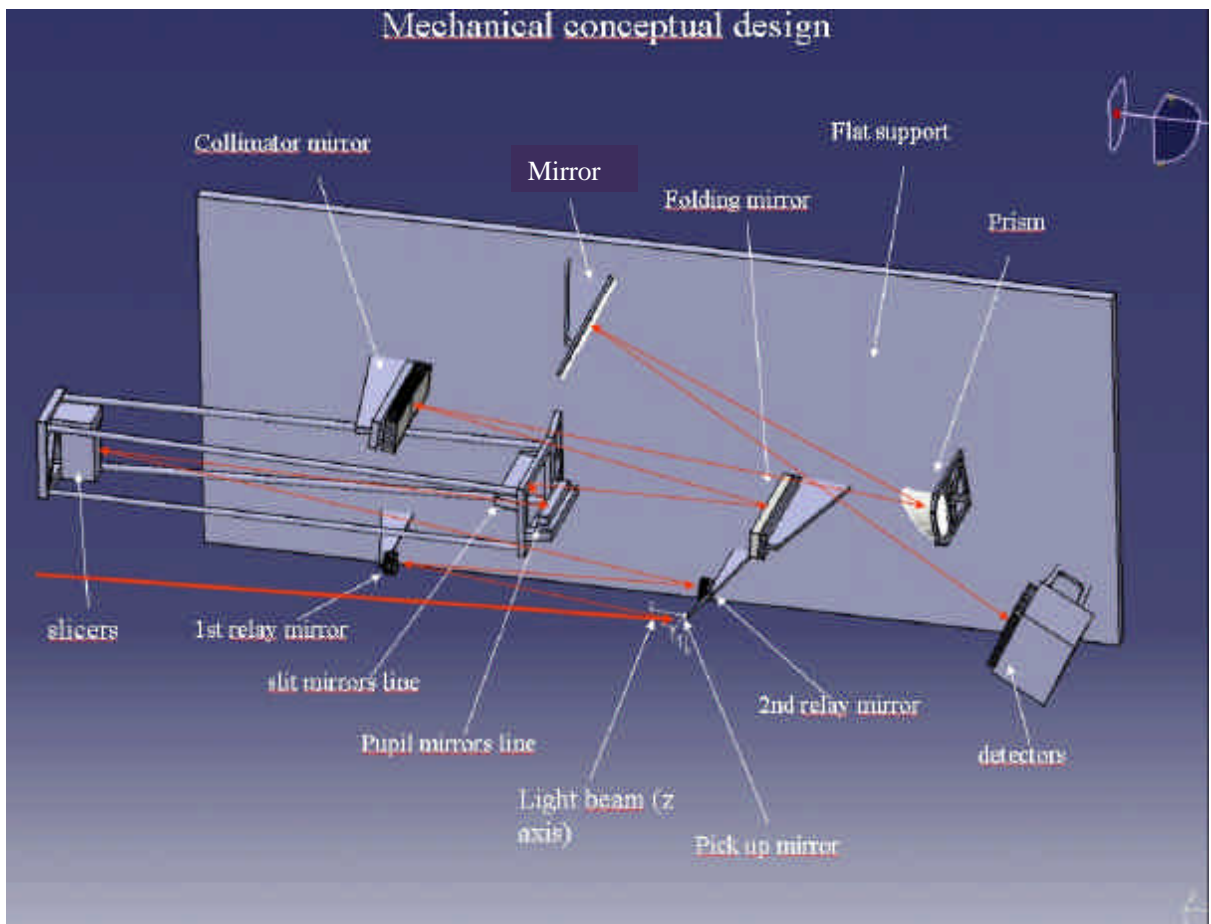

Figure 7: The global spectrograph instrument (only one channel is represented; the two channel version is under study).

**DETECTORS**

In the visible, the main goals are high quantum efficiency and very low noise. Given concerns over degradation due to radiation exposure and the poor performance of conventional thinned CCDs in the red part of the visible, we will study carefully the applicability of the LBNL CCDs. Thinned, backside-illuminated, low-noise conventional CCDs of 1024 × 1024 pixels are an alternative option.

In the IR, some factors constrain the detector technologies. The overall temperature for the SNAP instruments will be fixed in the range 130–140 K and the spectrograph must operate in this range. While keeping noise figures low, the cutoff wavelength of the array must be as close as possible to 1.7 µm. A 1024 × 1024 HgCdTe array with 18 µm pixels from Rockwell is under consideration. The choice of visible and IR detectors will be done in close collaboration with the teams designing the SNAP imager in order to maintain the simplest overall solution for SNAP. A detailed list of the performance specifications for the detectors is provided in Table 2. To achieve the listed performance in read noise and dark current, a multiple sampling technique is required. To optimize exposure time, the impact of the rate of cosmic rays on the readout noise is under present study.

|  | Visible | IR |
|---|---|---|
| Detector size | 1k × 1k | 1k × 1k |
| Pixel size | 15-20 µm | 18 µm |
| Detector temperature(K) | 140 | 140 |
| <QE>(%) | 80 | 60 |
| Read noise(e) | 2 | 5 |
| Dark current(e/pixel/s) | 0.001 | 0.02 |

Table 2. Spectrograph detector specifications

## 5. PRINCIPAL R&D EFFORT: THE SLICER UNIT

The proposed image slicer is of the same type as the one studied in the context of the NGST near-IR spectrograph (Allington-Smith *et al.*, 1999; Le Fèvre *et al.*, 1999). This technology has been ranked at NASA readiness level (TRL) 5 by a panel of NASA experts in the context of the concept appraisal of pre-phase A NGST studies. Readiness level 6 is required to be "space qualified." Prototyping activities are ongoing at Laboratoire d'Astrophysique de Marseille (LAM), in collaboration with other European institutes, to validate this technology both for large ground-based telescopes and for space applications, under funding by various agencies including ESA, CNRS and CNES. The R&D effort necessary to adapt this concept to the SNAP requirements therefore meshes nicely with ongoing activities and will be synchronized with the R&D phase. Specifically, we have an ongoing program to qualify image slicers for space instrumentation. We are now in the process of developing a realistic prototype for a space-qualified unit based on Zerodur-glass slices. Several slices have been manufactured successfully to specifications (Figure 6). This prototype adequately represents the SNAP application. The road map of the R&D is presented in Figure 8. This road map summarizes the European effort for image slicer development for space application.

ESA is funding a consortium lead by LAM and including Observatoire de Lyon, Durham University, and ESO, in order to push the slicer technology up to TRL 6. The development is ongoing for NGST and will fully validate this technique for the SNAP application.

Other contributions to this meeting provide details of different image slicer developments, design, performance studies, and other projects related to slicer technology [8,11].

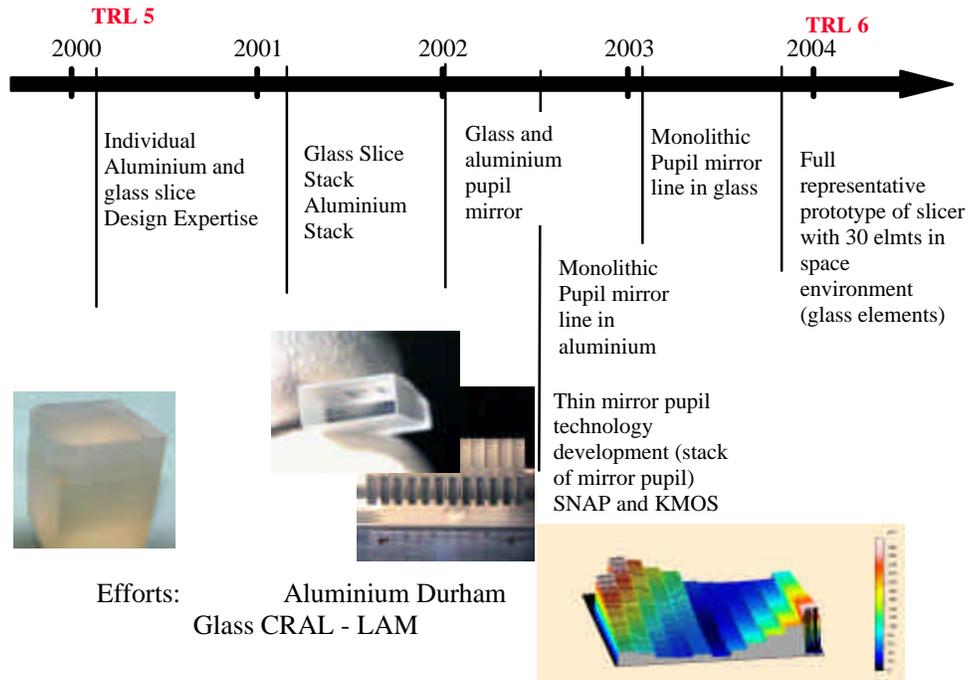

Figure 8: European slicer development road map.

## 6. EFFICIENCY ESTIMATE

Simulations of the efficiency of the instrument show very good results. Table 3 shows the cumulative efficiency of the instrument @ 0.9µm for visible and 1.1µm for the infrared. The first point to note is the conservative value assumed for the efficiency of individual mirrors; 98% is much lower than is expected from a silver coat at these wavelengths. The principal losses are from the prism (also conservatively estimated), and from the detector. Figure 9 shows the full response of the instrument for different detectors (LBL CCDs, HgCdTe with NGST specification, thinned HgCdTe with 60% of efficiency across the band, and WFC3 HgCdTe specs).

|  | telescope | Relay optics | Slicer Optic straylight diffra. | | | Spectro Mirrors prism dichroic | | | Detector Visible / NIR | |
|---|---|---|---|---|---|---|---|---|---|---|
| #elements | 4 | 3 | 3 | 1 | 1 | 3 | 1 | 1 | 1 | 1 |
| Efficiency/elt | 0.98 | 0.98 | 0.98 | 0.99 | 0.98 | 0.98 | 0.81 | 0.95 | 0.9 | 0.6 (0.8) |
| cumulative | 0.94 | 0.89 | 0.83 | 0.82 | 0.81 | 0.76 | 0.62 | 0.59 | 0.53 | 0.35 (0.47) |

Table 3: cumulative efficiency

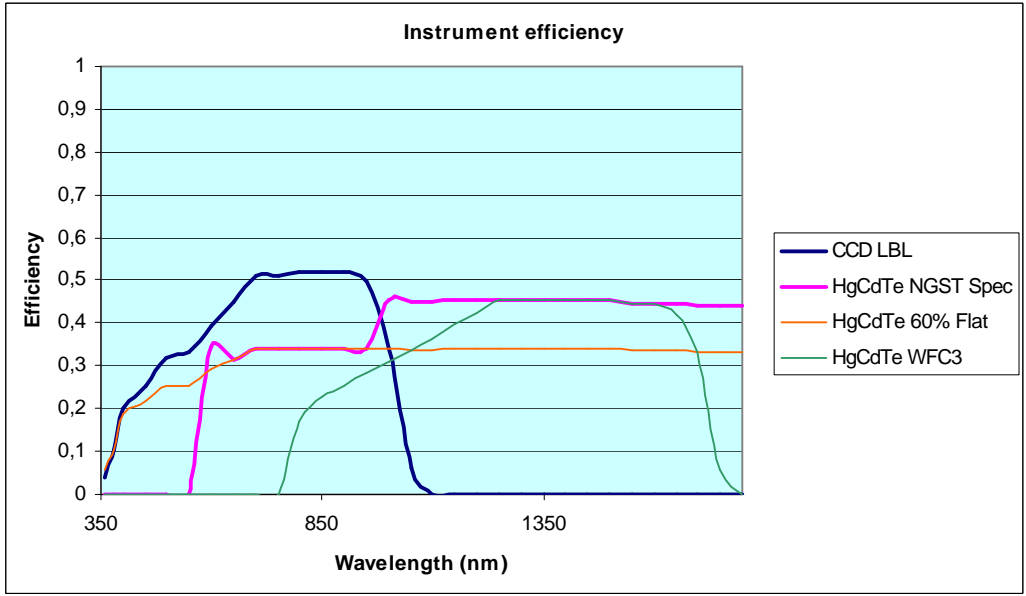

Figure 9: efficiency curve with different detector assumptions.

## 7. CONCLUSION

Spectral observations are one of the key programs of the SNAP mission. For a reasonable budget, volume, and risk, the proposed concept will achieve the objectives of the mission within the time allocation proposed in the mission strategy. Thanks to the European effort in the domain of image slicers, the project will benefit from the most efficient technique for 3D spectroscopy. The maturity of this technique is increasing and will be at Technical Readiness Level 6 (NASA scale) by the end of 2003. The SNAP project will benefit fully from this advance. The R&D effort for the spectrograph will be focused on the image slicer to qualify the small adaptation from the space-qualified prototype we are already building for ESA for NGST-NIRSPEC.